# Plasmon coupling in extended structures: Graphene superlattice nanoribbon arrays

**Daniel Rodrigo[a]\*, Tony Low[b], Damon B. Farmer[c], Hatice Altug[a] and Phaedon Avouris[c]\***

a. Institute of Bioengineering, École Polytechnique Fédérale de Lausanne (EPFL), Lausanne 1015, Switzerland.
b. Department of Electrical & Computer Engineering, University of Minnesota, Minneapolis, MN 55455, USA.
c. IBM Research Division, T.J. Watson Research Center, Yorktown Heights, NY 10598, USA.

**Abstract.** Interactions between localized plasmons in proximal nanostructures is a well-studied phenomenon. Here we explore plasmon-plasmon interactions in connected extended systems. Such systems can now be easily produced using graphene. Specifically we employ the finite element method to study such interactions in graphene nanoribbon arrays with a periodically modulated electrochemical potential or number of layers. We find a rich variation in the resulting plasmonic resonances depending on the dimensions and the electrochemical potentials (doping) of the nanoribbon segments and the involvement of transverse and longitudinal plasmon interactions. Unlike predictions based of the well-known "orbital hybridization model", the energies of the resulting hybrid plasmonic resonances of the extended system can lie between the energies of the plasmons of the individual components. The results demonstrate the wide range tunability of the graphene plasmons and can help to design structures with desired spectra, which can be used to enhance optical fields in the infrared region of the electromagnetic spectrum.

\* Corresponding authors: D.R.: daniel.rodrigo@epfl.ch; P.A.: avouris@us.ibm.com

Plasmonics is a well-developed and active field. It is based primarily on noble and coin metals, Au, Ag, and Cu. The excitation of localized plasmons in micro- and nanostructures made of these materials is used extensively to enhance linear and non-linear optical phenomena in the visible and near infrared sections of the electromagnetic (EM) spectrum [1-4]. Graphene is a semi-metal with properties that can further extend the plethora of phenomena and applications of plasmonics. Specifically, unlike conventional metals, the carrier density of graphene structures, and therefore their plasmonic resonances, are tunable by electrostatic or chemical doping over a wide range: $10^{11}$ to $10^{14}$ carriers/cm$^2$ [5-6]. Being a 2D material, graphene can easily be patterned using the well-developed planar patterning techniques of the semiconductor industry to generate new types of ultrathin and flexible plasmonic structures [7-10]. Due to the very high mobility of carriers in graphene, plasmons at low energies can be long-lived and have strong confinement factors ($\lambda_{vac}/\lambda_{graphene} \approx 100$) and large Purcell factors ($10^6$ to $10^7 \lambda_0^{-3}$) [11-14]. Finally, the resonance frequencies of graphene nano- and micro-structures cover the mid-infrared, far infrared, and terahertz sections of the spectrum and thus supplement and extend noble metal plasmonics [14-17].

Interactions between nearby localized plasmons of metallic particles and the resonances of complex shape structures are typically described very successfully by the 'orbital hybridization' model [2, 18-21]. In analogy with the interaction of atomic orbitals between two atoms that generate bonding and anti-bonding states, plasmonic resonances below and above the isolated resonances should result. This model has also been applied successfully to simple graphene plasmonic structures such as graphene rings [22] and dots [23]. However, the application of planar fabrication technology to graphene allows the generation of extended graphene structures with increased complexity. Understanding plasmonic interactions in such systems would allow the design of structures with prescribed plasmonic spectra. One such structure was recently studied experimentally [24]. It consists of a continuous single layer graphene nanoribbon array (GNRA) periodically overlaid with segments of a second graphene layer (Fig. 1a) [24]. It was found that application of the simple form of the orbital hybridization model could not account for the resulting plasmonic resonances. In this work we use this structure as an example to explore the nature of the plasmonic interactions in extended graphene structures typically referred to as a superlattice arrays [25]. We investigate first graphene nanoribbon arrays where the chemical potential is periodically modulated having alternating segments of respectively low and high chemical potential (Fig. 1b). Next, we explore graphene superlattice arrays where graphene nanoribbons alternate segments with different number of layers (Fig.1a). We show the connection between these two structures and analyze the nature of the new plasmonic modes supported by these novel graphene structures.

A typical extinction coefficient of the GNRA with alternating chemical potential is shown in Fig. 2a. We represent the extinction coefficient when the chemical potential of section 1 is kept constant ($\mu_{c1}$=0.35eV) and the potential of the section 2 is varied ($\mu_{c2}$). The structure parameters considered, unless stated otherwise, are a ribbon width and section length $W=L=100$nm, graphene chemical potential $\mu_{c1}$=0.35eV and relaxation time $\tau$=70fs. The graphene parameters are in the same range as those obtained from experimental data for nano-patterned CVD graphene [26]. The calculated resonance frequencies of GNRAs with uniform chemical potential $\mu_{c1}$ and $\mu_{c2}$ are represented with respective dashed lines that follow the theoretical dependence $\omega_{SPP} \sim \sqrt{\mu_c}$ [7] and are in excellent agreement with analytically calculated resonance frequencies (Fig. S1). We observe that the single resonance of the uniform GNRA ($\mu_{c2}=\mu_{c1}$) is transformed into a multi-resonance response when $\mu_{c2} \neq \mu_{c1}$. The lower energy mode is spectrally located between the resonances of $\mu_{c1}$ and $\mu_{c2}$ uniform-potential GNRAs (dashed lines). The second mode is located either below or above the $\mu_{c2}$ resonance depending on the parameters and in some cases, a third mode can appear (for instance for $\mu_{c2}=3\mu_{c1}$). This response cannot be described using the orbital hybridization model (Fig. 2b), which predicts bonding and anti-bonding modes producing resonances respectively located below and above the individual resonance frequencies for $\mu_{c1}$ and $\mu_{c2}$. However, the orbital hybridization model can successfully describe the case in which each nanoribbon has uniform potential and adjacent nanoribbons in close proximity alternate between $\mu_{c1}$ and $\mu_{c2}$ (Fig. 2c). While this second structure shows clearly a bonding and anti-bonding mode, further description beyond orbital hybridization is needed for the nanoribbon superlattice.

The nature of the multi-section GNRA is strongly determined by the length $L$ of its sections. When sections are much shorter than the plasmon wavelength ($L \ll \lambda_{SPP}$ or equivalently L≪W) variations in the potential are subsequently produced at a much smaller scale than the plasmon wavelength and as a result, graphene behaves as a uniform effective medium. These short-section structures therefore operate as uniform-potential GNRAs with an average chemical potential between $\mu_{c2}$ and $\mu_{c1}$, producing a single resonance located between those of uniform GNRAs with $\mu_{c2}$ and $\mu_{c1}$ respectively (Fig. 3a). On the other hand, when sections are much longer than the plasmon wavelength ($L \gg \lambda_{SPP}$ or equivalently L≫W), the two sections support independent plasmonic resonances. As a result, the long-section structures behave as two independent GNRAs with uniform potential $\mu_{c1}$ and $\mu_{c2}$ producing two separate resonances matching those of the uniform GNRAs (Fig. 3b). The transition from short to long sections is explored in Fig. 3c for a varying length $L$ and $\mu_{c2}=2\mu_{c1}$. While the two extreme cases for the section length ($L \ll \lambda_{SPP}$ and $L \gg \lambda_{SPP}$) produce plasmonic behavior of uniform-potential GNRAs, the transition cases between short and long sections ($L \approx \lambda_{SPP}$) show additional resonance peaks that require further study.

To understand the nature of the new modes arising when the section length is comparable to the plasmon wavelength we study the charge distribution and dispersion of these modes for a varying plasmonic wavevector by changing the ribbon width $W$ (Fig. 4a). The dashed lines show the simulated resonance frequencies of uniform-potential nanoribbons at $\mu_{c1}$ and $\mu_{c2}$, which follow the theoretical dispersion for graphene plasmons $\omega_{SPP} \sim 1/\sqrt{W}$ (Fig. S1) [7].

We see that the interaction leads to two branches in the dispersion curve of the system (Fig. 4a). The low frequency mode labeled *m0* is continuous and its energy lies overall above the energy of the single layer energy dispersion (lower dashed curve). The other branch is discontinuous showing a number of gaps. Some points, labeled *m*1, *m*2, *m*3, ... lie on top the dispersion of the two-layer graphene (upper dashed line), while others lie below (m$^-$) or above (m$^+$) that dispersion curve. To understand the physical origin of this behavior we plot the electric charge distribution for each mode (Fig. 4b). We observe that the field over section 2 (double layer section) represents a charge separation (dipole formation) as expected from the excitation of the fundamental plasmonic mode, although variations in the spatial distribution of charges can be seen. Over section 1 (monolayer section) an increasing number of field oscillations is observed. The origin of the oscillations can be understood by considering the radii of the plasmon iso-energy surfaces of regions 1 and 2. Region 1 has a lower chemical potential than region 2 ($\mu_{c1}<\mu_{c2}$). Now the plasmon wavevector in graphene is given by $k \propto (\hbar\omega)^2/\mu \propto (\hbar\omega)^2/\sqrt{\eta}$ [7, 10]. Thus the iso-energy surface radius of section 1 is larger than that of section 2, i.e. it contains a larger number of wavevectors. In the higher doped section 2, the transverse momentum imparted by the finite ribbon width, $q_y \approx 3\pi/4W$ [27], lies along the iso-energy contour and is excited when it is in resonance with the incident light. The longitudinal component of the wavevector in the low density section 1 reduces its transverse component [28] and helps to match it to the wavevector in section 2. Plasmon waves with the extra wavevectors in section 1 cannot propagate 1 into section 2. These waves are then reflected at the boundary and form standing waves within region 1 akin to Fabry-Perot type oscillations. Depending on the size of section 1, an even or odd number of plasmon half-wavelengths can be accommodated. When an even number fits in section 1, then the dipole moment of this segment vanishes, there is minimal interaction between segments, the energies of modes *m*1, *m*2, *m*3, ... lie on top of the double layer dispersion and the charge distribution becomes uniform along section 2. On the other hand, when an odd number of half wave lengths is present, then a dipole is formed in region 1, which can be parallel or anti-parallel to that of the dipole of section 2, thus leading to modes higher (*m*$^+$) or lower in energy (*m*$^-$), respectively (Fig. 4c). Since the dipole of section 1 flips its direction when the mode number increases by one, this results in the formation of gaps in the dispersion.

In contrast with the higher order modes, the *m0* mode does not show any nodes over section 1. The electrical charge in mode *m0* is mainly located over section 1 except for the lowest wavevectors, whose charge is equally distributed over sections 1 and 2. As a consequence, the resonance of mode *m0* overlaps that of a $\mu_{c1}$ uniform-potential GNRA, except for the lowest wavevectors whose resonance is located between $\mu_{c1}$ and $\mu_{c2}$ resonances (Fig. S2). We also note that that these results are not significantly modified when we allow for a smooth transition between $\mu_{c1}$ and $\mu_{c2}$ (Fig. S3). This indicates that the new modes supported by the alternating potential GNRA are created by the periodic modulation of the chemical potential rather than by an abrupt discontinuity between sections.

To further understand the physics of the alternating potential GNRA we extend the dispersion analysis to different chemical potentials $\mu_{c2}$. We observe in Fig. 5a that for the different values of $\mu_{c2}$, the modes *m1*, *m2*, and higher are rescaled along the horizontal axis to follow the shift of the resonance corresponding to a uniform $\mu_{c2}$ GNRA (top dashed line). On the other hand, there is no variation in the range of allowed frequencies for these modes (i.e. no variation in the vertical scale). Indeed, these allowed frequencies are controlled by the standing wave created along section 1 and in particular they depend on the section length $L$ as shown in Fig. 5b. As the section length $L$ is increased, the modes *m1*, *m2*, and higher are excited at lower frequencies. This is evidenced in Fig. 5b as a vertical rescaling along the frequency axis. We observe that for short lengths only mode *m0* is active and is located on between the resonances of $\mu_{c1}$ and $\mu_{c2}$ uniform-potential GNRAs (dashed lines). As $L$ is progressively increased modes *m1*, *m2*, and above are excited at lower frequencies and provide additional resonance peaks. Finally, for long lengths, we see that modes *m1*, *m2*, and higher group together along the $\mu_{c2}$ resonance dashed line, while mode *m0* overlaps with the $\mu_{c1}$ resonance dashed line. These observations provide an explanation for the multiple peaks shown first in Fig. 2 and give a complete description for the transition between the single-resonance of short sections ($L \ll \lambda_{SPP}$) and the double-resonance of long sections ($L \gg \lambda_{SPP}$) initially shown in Fig. 3.

We investigate next the graphene superlattice array where nanoribbons are divided in sections alternating between single-layer and double-layer graphene. The double-layer region is composed of two identical layer separated by an interlayer distance $g$. The extinction spectra of the GNRA for different values of $g$ are shown in Fig. 6a and 6b. For the smallest gap ($g=W/300$) we observe that the extinction spectrum is in excellent agreement with that of the alternating chemical potential model ($\mu_{c1}$-$\mu_{c2}$). This due to the strong coupling between the plasmons in the two layers when $g \ll \lambda_{SPP}$ [16]. More generally, a stack of graphene layers separated by infinitesimal spacing ($g=0$) is electromagnetically equivalent to a single layer incorporating the conductivities of the individual layers. Since the added conductivity of parallel graphene layers with Drude conductivity is

$$\sigma_{eq} = \sum_j \sigma_j = \sum_j \frac{i\,e^2}{\hbar\pi} \frac{\mu_c^{(j)}}{\omega + i/\tau}$$

[29], where $\mu_c^{(j)}$ is the chemical potential of each layer, we can replace multi-layer graphene by single-layer graphene with equivalent chemical potential $\mu_{c,eq} = \sum_j \mu_c^{(j)}$. As a result, the structures with alternating chemical potential and alternating number of layers are equivalent when the interlayer distance $g$ tends to zero.

As the gap $g$ increases, the coupling between layers weakens and we observe additional plasmonic resonances (*s1*, *s2*, *s3*, etc. in Fig. 6b) that were not initially observed using the alternating chemical potential model. The origin of these resonances can be determined by observing their corresponding charge density distributions (Fig. 6c). The charge density for modes *m0* and *m1* shows a high-order Fabry-Perot resonance coupled over the double-layer section (section 2). When the charges on the top and bottom layers are added the high-order Fabry-Perot oscillation vanishes and what is left is a non-zero charge distribution that is identical to that obtained from the $\mu_{c1}$-$\mu_{c2}$ model (Fig. S4). The multiple orders of the high-order Fabry-Perot oscillations on the double-layer section create the additional resonances *s1*, *s2*, etc. that were not present in the single-layer model. Finally, when the gap distance becomes comparable to or larger than the plasmon wavelength (e.g. *g=W*) the two layers become uncoupled generating two resonances ($m_{\text{bottom}}$ and $m_{\text{top}}$) independently produced by the bottom and top layers. In summary, the multi-layer superlattice is equivalent to the alternating potential structure for *g=0*, due to the strong coupling between plasmons in the two layers. This coupling becomes progressively weakened as *g* increases and the plasmonic response of the multi-layer superlattice converges towards that of two isolated graphene layers.

The results presented above demonstrate that graphene superlattice nanostructures that combine multiple number of layers provide great versatility to tailor the plasmonic response beyond that of canonical structures such as nanoribbons or nanodots. The simulations reveal the nature of the new plasmonic modes arising in these superlattice arrays and provide a guide to adjust the number of resonances in the structure, as well as their spectral position. As our ability to fabricate increasingly complex multilayer devices improves and the number of available two-dimensional materials increases, developing our understanding of complex plasmonic modes will ultimately enable us to engineer the plasmonic resonance response of two-dimensional nanostructures.

**Computational details.** The simulations are carried out calculating the full 3D electromagnetic field in the structure by using the finite elements method (Ansys HFSS v15 software) to solve Maxwell equations in the frequency-domain. The simulated structure consists of a three-dimensional unit cell arranged in a two-dimensional periodic array. The electromagnetic excitation consists of a plane wave with orthogonal incidence angle and polarization transverse to the graphene nanoribbons. Graphene is modeled as a zero-thickness two-dimensional surface enforcing an impedance boundary condition where the ratio between tangential components of the electric and magnetic fields is equal to graphene conductivity. Graphene surface conductivity is calculated using the Drude model as

$$\sigma = \frac{i\,e^2}{\hbar\pi} \frac{\mu_c}{\omega + i/\tau}$$

[29], including the temperature effect (*T*=300ºK) as

$$\mu_c = \mu_{c,0} + 2\,k_B T \ln\left(1 + e^{-\frac{\mu_{c,0}}{k_B T}}\right)$$

The relaxation time for nano-patterned CVD graphene is $\tau$=70fs in accordance with experimental data [26]. Graphene is supported by a dielectric substrate with refractive index $n_s$=1.4 and the interlayer volume in the double-layer model has refractive index $n_{2L}$=1. The structure periodicity is defined by two pairs of Floquet-Bloch periodic boundary conditions on the surfaces delimiting the unit cell. We use an initial mesh with tetrahedral elements having a maximum length of $W/15$. The mesh is iteratively refined by increasing the number of mesh elements by 30% each iteration. The convergence criterion is defined as $|t^N - t^{N-1}| < 0.02$, where $t$ is the complex transmission coefficient and N is the iteration number. The electric surface charge density ($q_s$) is calculated from the charge conservation equation as $q_s = -\nabla J_s$, where $J_s$ is the surface current in graphene. This charge density is validated against that obtained from Gauss' law $q_s \sim E_z(z = 0^+) - E_z(z = 0^-)$ showing good agreement (Fig. S5). Experimental validation of the computational technique for graphene nanoribbon arrays can be found in ref. [17].

**Supporting information.** Analytic calculation of GNRA resonant frequencies, smooth transition between sections, charge density distribution in mode *m0*, comparison between charge density in single- and multi-layer GNRAs, charge distribution calculation from Gauss law and charge conservation equation.

**Notes.** The authors declare no competing financial interest

**Acknowledgments.** This work was funded in part by European Commission (FP7-IEF-2013-625673-GRYPHON), the Swiss National Science Foundation (SNSF) through project 133583.

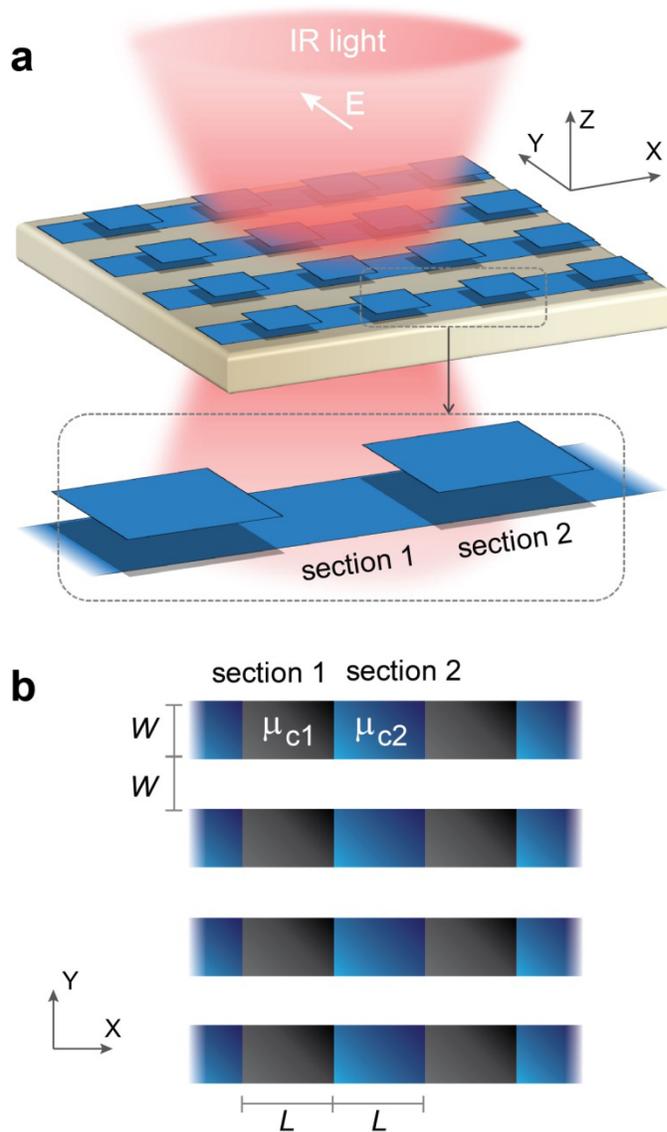

**Figure 1**. (a) Conceptual view of the graphene nanoribbon array (GNRA) with alternating number of layers. Each nanoribbon alternates sections of single-layer (section 1) and double-layer graphene (section 2). The plasmonic excitation is performed by infrared light illumination polarized transversally to the nanoribbons. (b) GNRA with modulated chemical potential. Each nanoribbon alternates sections of low chemical potential ($\mu_{c1}$ in section 1) and high chemical potential ($\mu_{c2}$ in section 2).

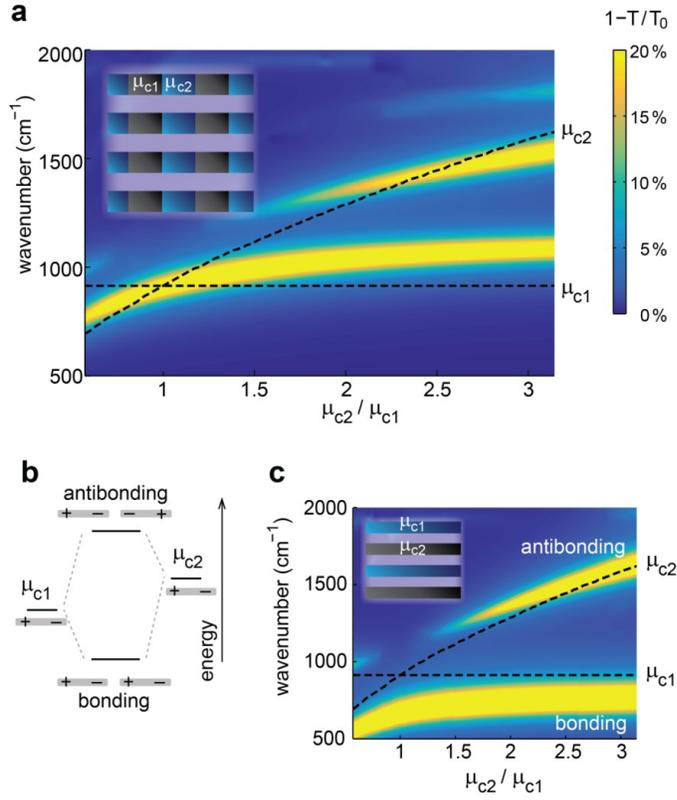

**Figure 2.** (a) Extinction coefficient (1-$T/T_0$) of the GNRA with alternating chemical potential ($W=L=$100nm, $\mu_{c1}=$0.35eV, $\tau=$70fs). T and $T_0$ are the transmission coefficients with and without graphene nanoribbons, respectively. Dashed lines represent the resonance frequency of GNRAs with uniform-potential $\mu_{c1}$ and $\mu_{c2}$. (b) Schematic description of the orbital hybridization model, showing the energies of the bonding and anti-bonding hybridized modes. (c) Extinction coefficient of a GNRA with uniform nanoribbons where adjacent ribbons alternate between $\mu_{c1}$ and $\mu_{c2}$.

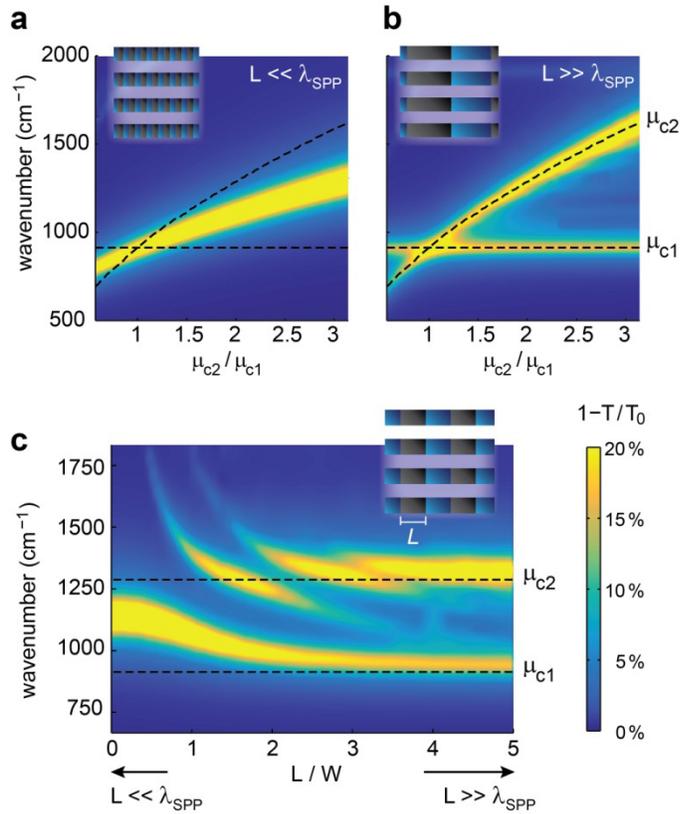

**Figure 3.** (a) Extinction coefficient of the GNRA with alternating chemical potential for a section length $L$ much shorter than the plasmon wavelength ($L=W/5=20$nm) and (b) much longer than the plasmon wavelength ($L=5W=500$nm). Dashed lines represent the resonance frequency of uniform-potential GNRAs. (c) Extinction for a section length $L$ covering the transition between short and long sections ($\mu_{c2}= 2\mu_{c1}=0.7$eV).

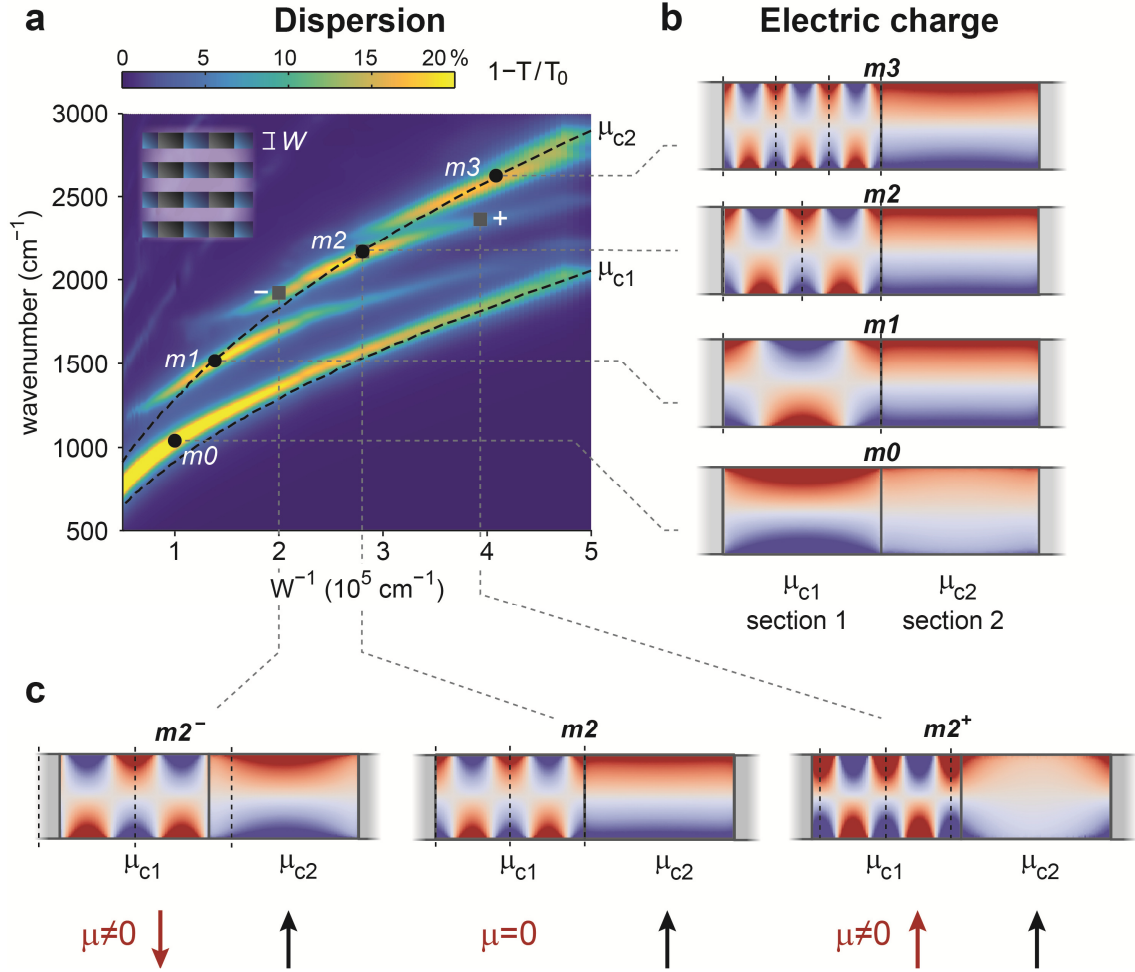

**Figure 4.** (a) Dispersion of the plasmonic modes supported by the alternating chemical potential GNRA ($L$=100nm, $\mu_{c2}$= $2\mu_{c1}$=0.7eV). (b) Electrical charge density distribution for each mode. The corresponding frequency and wavevector for each charge distribution are indicated with corresponding markers in (a). The red/blue color indicates the positive/negative sign of the charge density. (c) Electrical charge density distribution for mode *m2* at different positions along the dispersion curve.

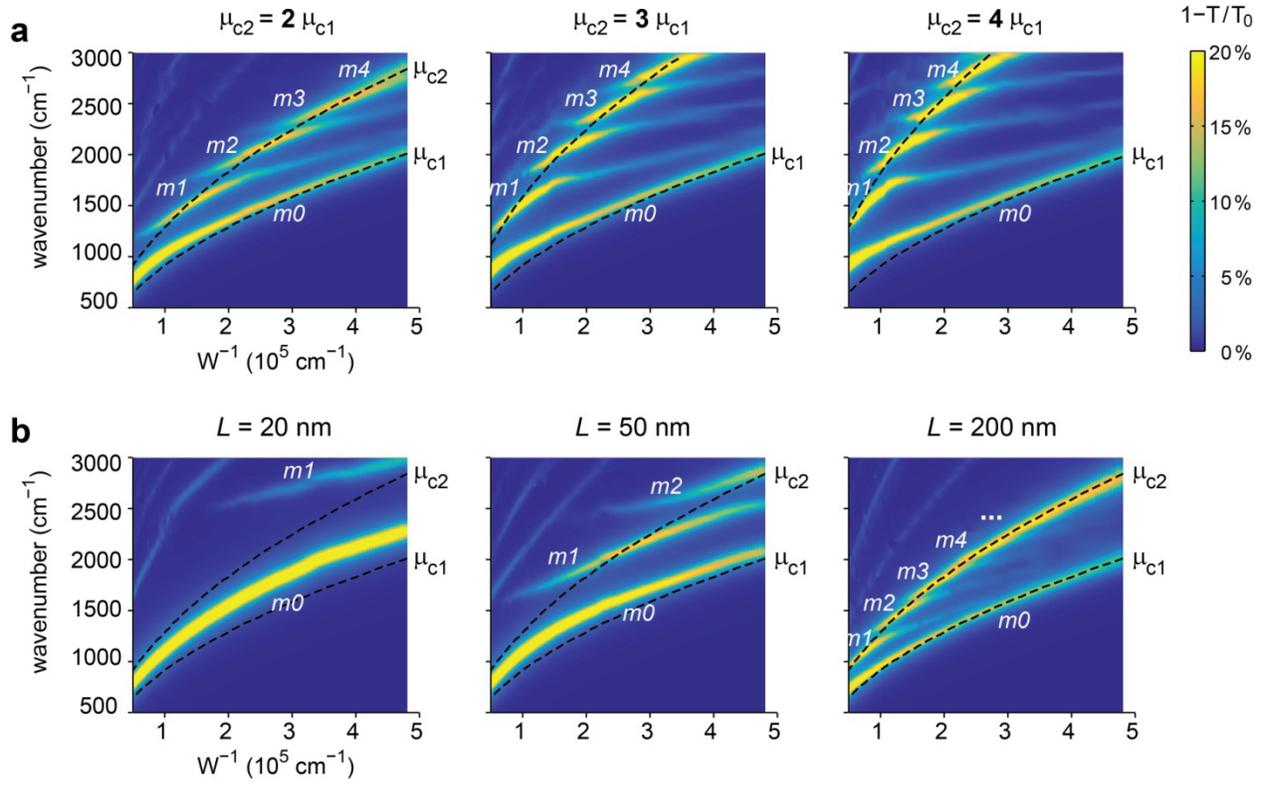

**Figure 5.** Dispersion of the plasmonic modes of the alternating chemical potential GNRA for (a) increasing values of chemical potential in section 2 and (b) different section lengths *L*.

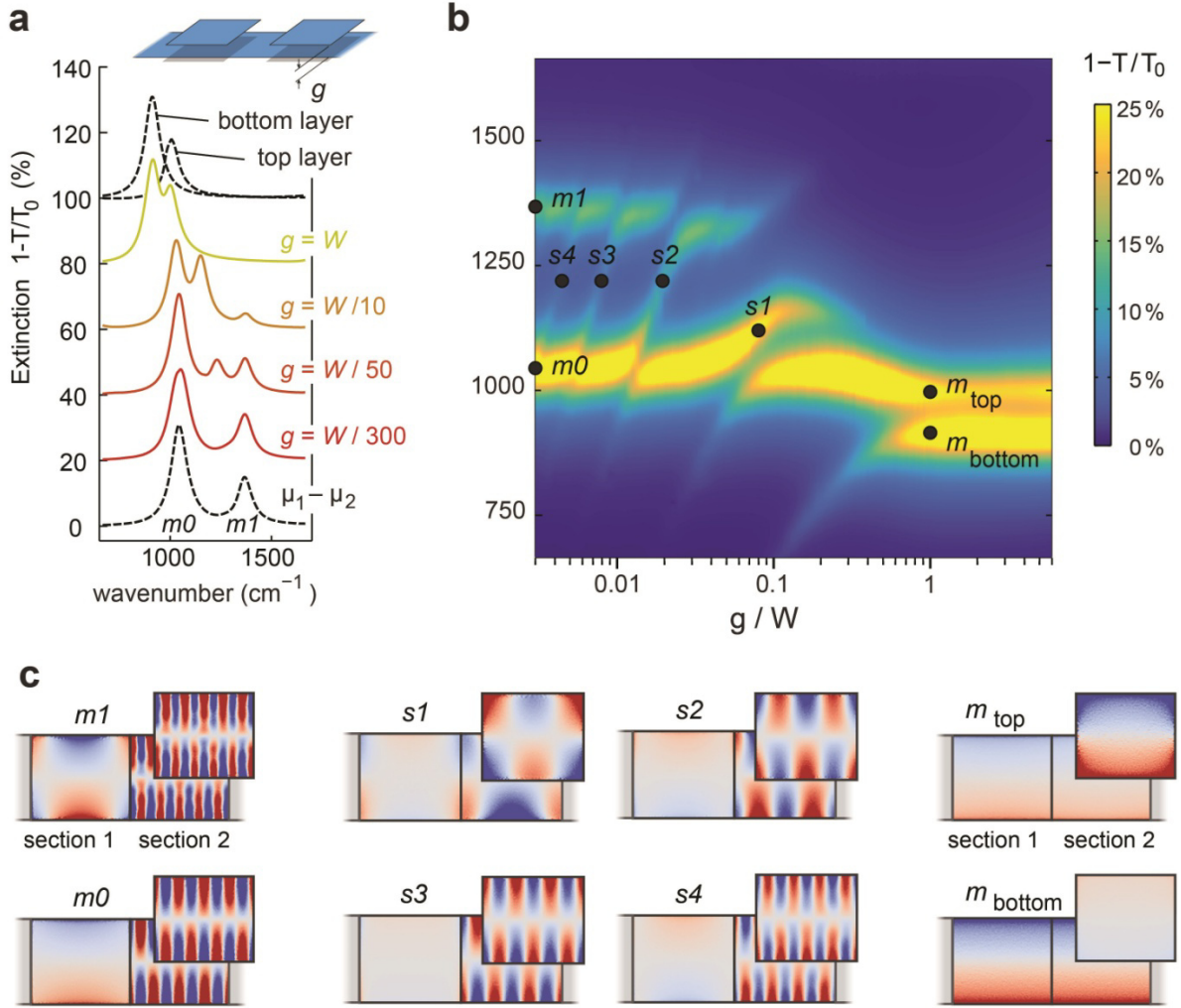

**Figure 6.** (a,b) Extinction spectrum of the GNRA with alternating single-layer and double-layer graphene sections ($W=L=100$nm,). The two-layer section is considered as two identical graphene layers ($\mu_c=0.35$eV) separated by an air gap with thickness $g$. Dashed lines correspond to the spectrum of the alternating chemical potential model ($\mu_{c2}=2\mu_{c1}=0.7$eV) and isolated top/bottom layers. (c) Electrical charge density distribution for each mode at the frequency and gap $g$ indicated in (b).

# Plasmon coupling in extended structures: Graphene superlattice nanoribbon arrays

**Daniel Rodrigo[a]\*, Tony Low[b], Damon B. Farmer[c], Hatice Altug[a] and Phaedon Avouris[c]\***

a. Institute of Electrical Engineering, École Polytechnique Fédérale de Lausanne (EPFL), Lausanne 1015, Switzerland.
b. Department of Electrical & Computer Engineering, University of Minnesota, Minneapolis, MN 55455, USA.
c. IBM T.J. Watson Research Center, Yorktown Heights, NY 10598, USA.

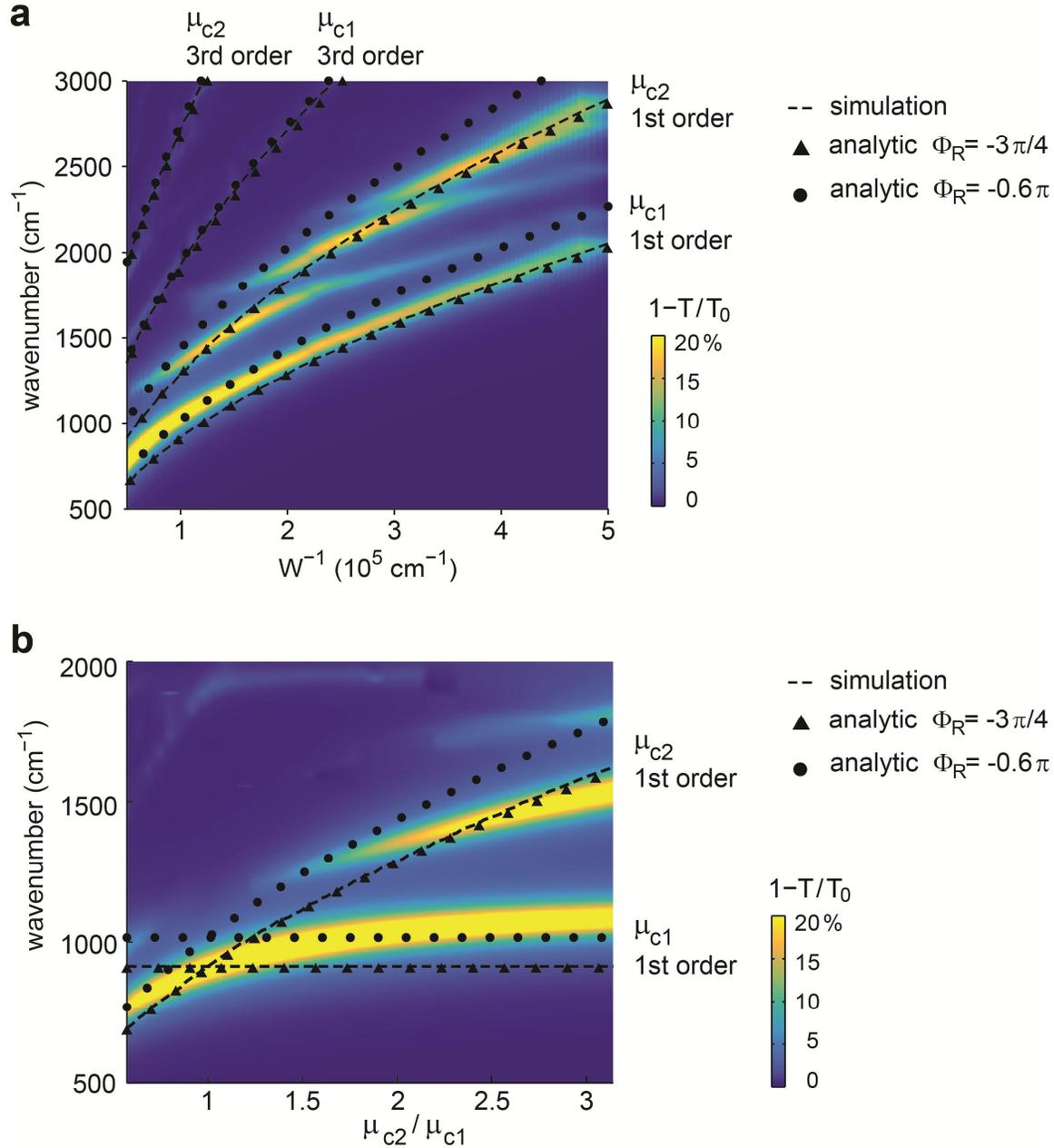

**Figure S1.** Comparison between simulated and analytically calculated resonance frequencies for uniform GNRA for varying (a) width $W$ and (b) chemical potential $\mu_c$. Resonance frequencies are overlaid over the simulated extinction of an alternating chemical potential GNRAs corresponding to figures 2a and 4a. Analytic calculations of resonance frequency are performed as $\omega^2 = \frac{e^2}{2\pi\,\varepsilon_0\varepsilon_r\,\hbar^2}(\pi n - \pi - \phi_R)\,\mu_c/W$, where $n$ is the order of the mode and $\phi_R$ is the reflection phase at the graphene edge [27]. There is excellent agreement between the simulated and analytically calculated curves for $\phi_R = -0.6\,\pi$.

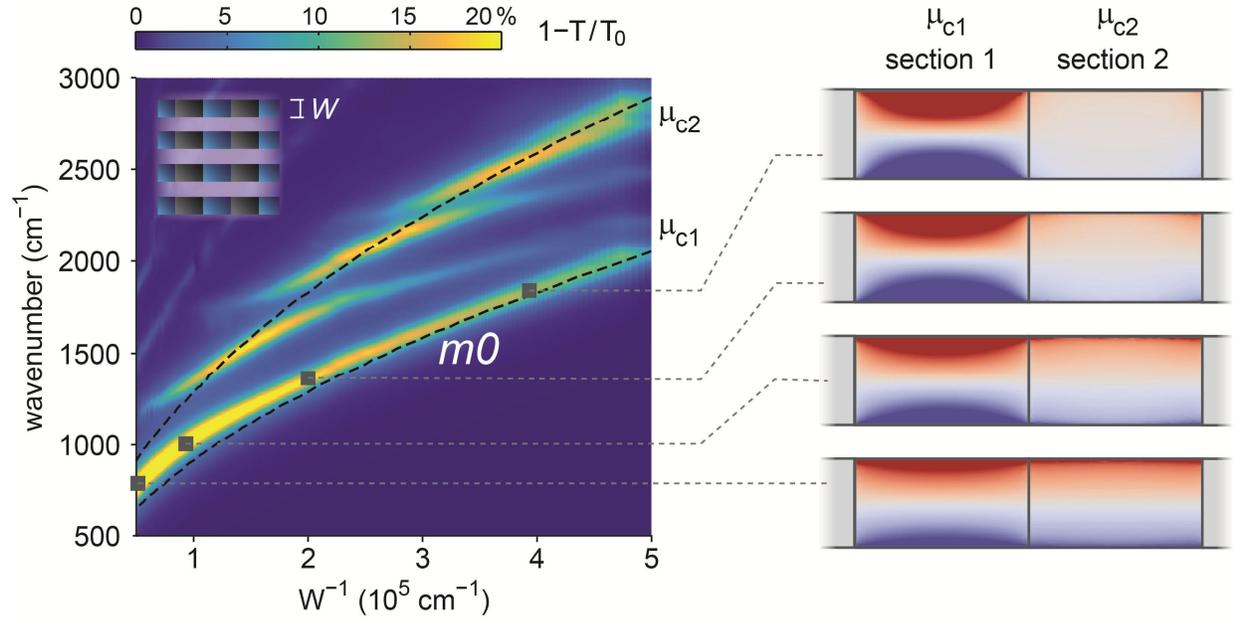

**Figure S2.** Dispersion and charge density distribution for mode *m0*. For highest wavevectors the charge density is localized on section 1 and the resonance frequency matches that of $\mu_{c1}$ uniform-potential GNRA. For lowest wavenumbers the charge density is spread over sections 1 and 2, and the resonance frequency is between the resonances of $\mu_{c1}$ and $\mu_{c2}$ uniform-potential GNRA.

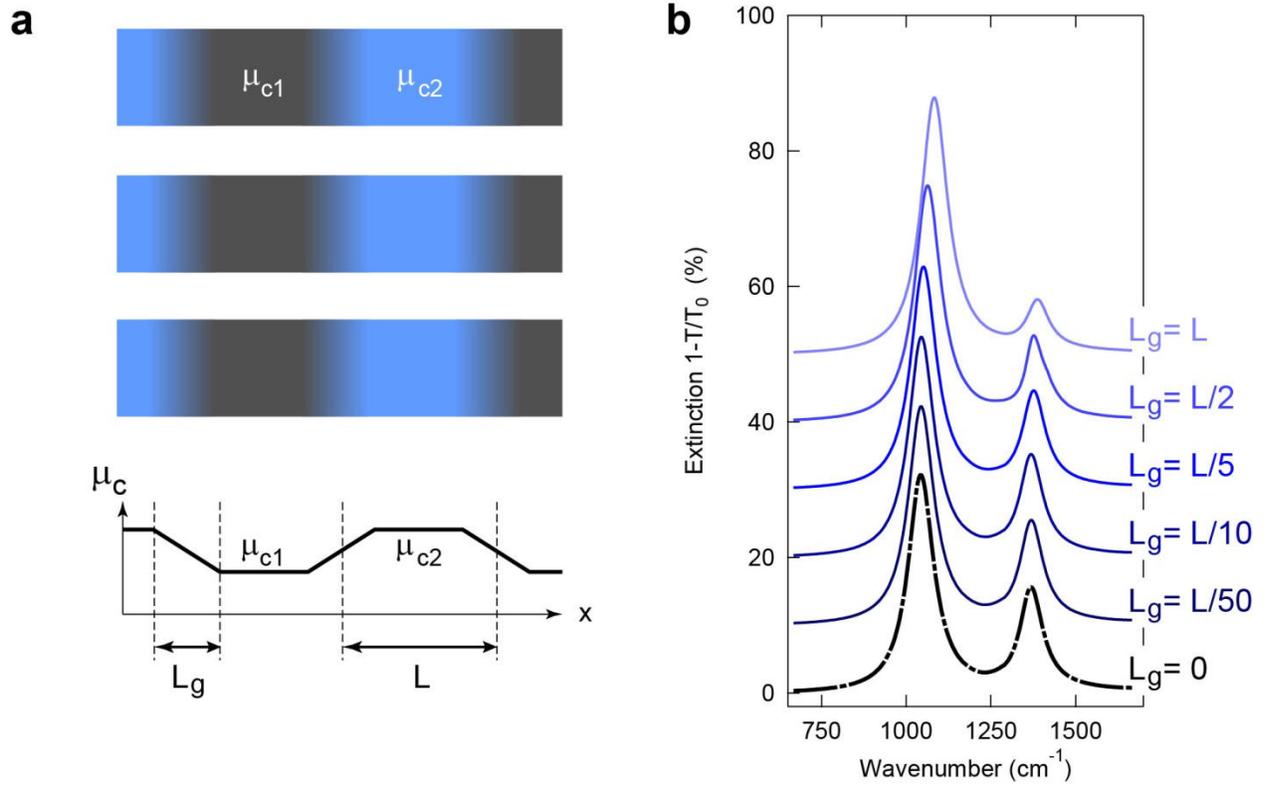

**Figure S3.** (a) Alternating chemical potential GNRA with smooth transition between sections. The chemical potential is linearly tapered over a transition region of length $L_t$. (b) Extinction spectra for different transition length $L_t$. The spectral response of the alternating potential GNRA is preserved even for transition lengths $L_t$ comparable to the section length $L$, showing that the new modes supported by the alternating potential GNRA are created by the periodic modulation of the chemical potential rather than the by abrupt discontinuity between sections.

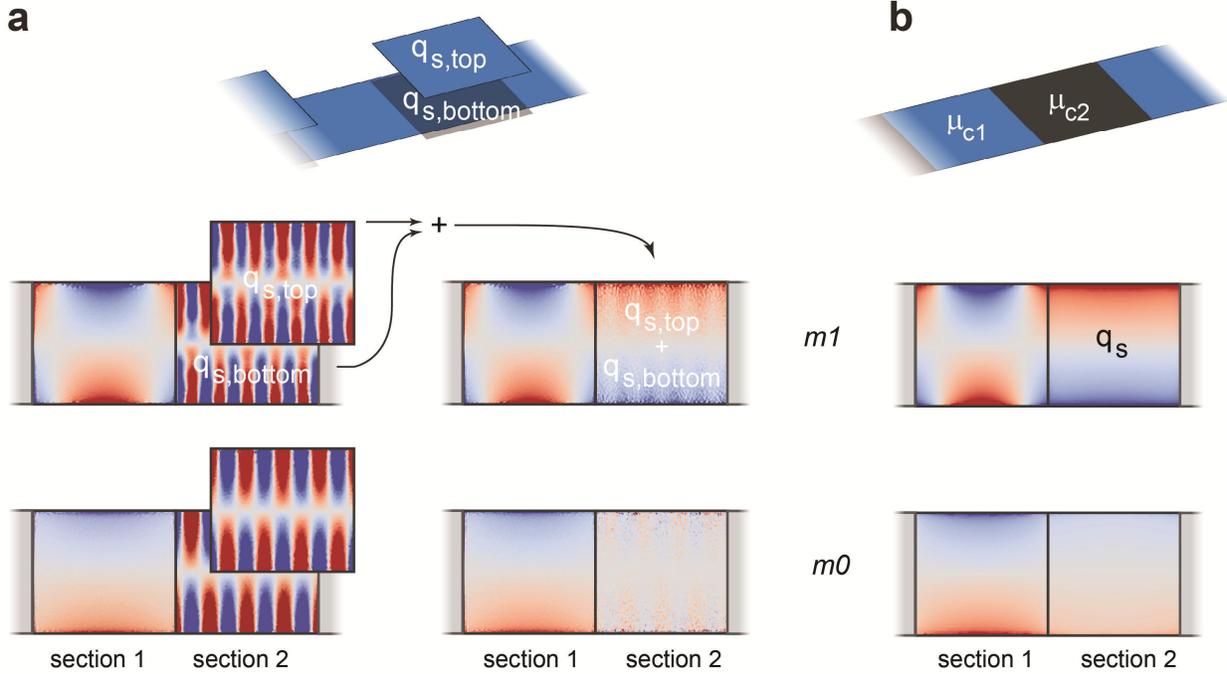

**Figure S4.** (a) Charge density distribution in the alternating layer model ($W=L=100$nm, $\mu_c=0.35$eV) for modes $m1$ and $m0$. The charge of the two layer section is firstly depicted individually for each layer and secondly the addition of charge density from the top and bottom layers is shown. (b) Charge density distribution in the equivalent alternating chemical potential model ($W=L=100$nm, $\mu_{c2}=2\mu_{c1}=0.35$eV). The calculated charge density is in good agreement by that resulting from adding the charges in the top and bottom layers in the multi-layer model.

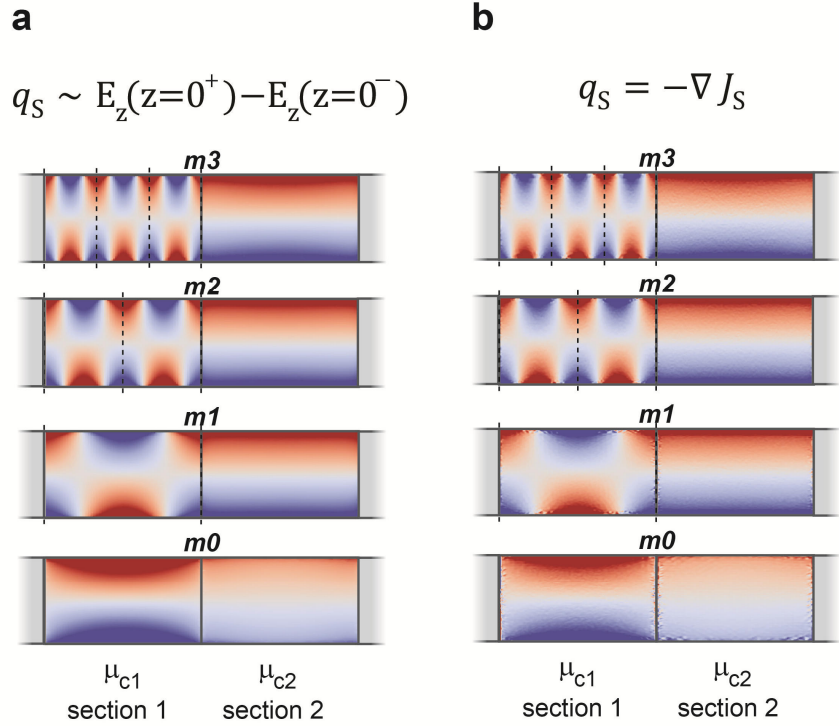

**Figure S5.** Charge density distribution for modes m0, m1, m2 and m3 calculated from (a) electric field $E$ using Gauss law and (b) surface current density ($J_s$) using the charge conservation law. Excellent agreement between the two methods is obtained.

6